\documentclass{ar-1col-S2O}

    \usepackage{url}
    \usepackage[linktocpage, colorlinks=true, allcolors=blue, breaklinks=true]{hyperref}
    \usepackage{siunitx}

    \usepackage[
    backend=biber,
    isbn=false,
    url=false,
    doi=true,
    style=nature, %
    maxbibnames=99, 
    sorting=none,
    backref=true
    ]{biblatex}
    \DeclareFieldFormat[article]{titlecase}{\MakeSentenceCase*{#1}}

    \jname{Annu. Rev. Cond. Matt. Phys.}
    \jvol{AA}
    \jyear{2026}
    \doi{10.1146/((please add article doi))}

\begin{document}

    \markboth{Mathijssen et al.}{Biomedical Active Matter}
    \title{Biomedical active matter: Emergence and breakdown of collective functionalities}
 
    \author{%
    Arnold J. T. M. Mathijssen,$^{1,\dag}$
    Hamed Almohammadi,$^{1,*}$
    Lauren Altman,$^{1,*}$ 
    Talia Calazans,$^{1,*}$ 
    M. J. Ferencz,$^{1,*}$ 
    Michelle Fung,$^{1,*}$ 
    Ian J. Lee,$^{1,*}$ 
    Maciej Lisicki,$^{1,2,*}$ 
    Ivy Liu,$^{1,*}$ 
    Maggie Liu,$^{1,*}$
    Tianyi Liu,$^{1,*}$
    Ernest Park,$^{1,*}$ 
    Ran Tao,$^{1,*}$ 
    Albane Th{\'e}ry,$^{1,3,*}$
    Zeyuan Wang,$^{1,*}$
    and
    Margot Young.$^{1,*}$
    \affil{$^1$University of Pennsylvania, Philadelphia, PA 19104, United States}
    \affil{$^2$University of Warsaw, Pasteura 5, 02-093 Warsaw, Poland}
    \affil{$^3$University of Warwick, Coventry, CV4 7AL, United Kingdom}
    \affil{$^\dag$Leading and corresponding author, amaths@upenn.edu}    \affil{$^*$These authors contributed equally and are listed alphabetically.}
    }

    \begin{abstract}
    Living systems are made of active materials with microscopic components that work together to perform macroscopic biological tasks. The breakdown of these collective functionalities leads to diseases, which, conversely, could be treated by exploiting self-organization in healthcare technologies. Here, we review recent advances in this rapidly growing field of biomedical active matter. The main themes are (1) collective self-assembly and spatiotemporal coordination; (2) collective motion, transport, and navigation; (3) collective sensing, signaling, and communication; and (4) collective adaptation, evolution, and learning. We discuss these emerging processes in a wide range of systems, including protein folding, biomolecular condensates, cytoskeleton dynamics, intracellular flows, bacterial biofilms, quorum sensing, cilia synchronization, wound healing, biolocomotion, neurons, endocrine signalling, and cardiovascular flow networks. For each, we highlight medical conditions associated with reduced collective functionality and how they may be treated using microrobotic swarms, bioinspired metamaterials, diagnostics, lab-on-chip devices, organoids, and other active and adaptive matter innovations.  
    \end{abstract}


    \maketitle
    \tableofcontents


\section{INTRODUCTION}
\label{sec:Introduction}

Together, we can do more than alone.
Indeed, cooperation is found everywhere in biology: 
Molecules construct cells, cells form tissues, tissues develop into animals, and animals work together in groups.
This upward cascade is supported by collective functionalities, including collective sensing, actuation, assembly, information processing, adaptation, and learning.
But how do these collective functionalities emerge?

The field of active matter \cite{Vicsek1995NovelParticles, Toner1995Long-RangeTogether, Gompper2025TheRoadmap} aims to establish fundamental principles that govern how microscopic interactions can lead to macroscopic phenomena. 
Natural active matter can reconstitute biological mechanisms from only the essential living parts \cite{Needleman2017ActiveBiology, Aranson2022BacterialMatter}, while synthetic active matter can reproduce life-like behaviors with entirely non-living components \cite{Bechinger2016ActiveEnvironments, Zottl2016EmergentColloids}.
This field is truly interdisciplinary, bridging physics with 
cell biology \cite{Needleman2017ActiveBiology}, 
chemistry \cite{Dauchot2019ChemicalMatter}, 
nanotechnology \cite{Nelson2010MicrorobotsMedicine}, 
hydrodynamics \cite{Marchetti2013HydrodynamicsMatter, Lauga2020TheMotility}, 
statistical mechanics \cite{Ramaswamy2010TheMatter}, 
bacteriology \cite{Aranson2022BacterialMatter}, 
food science \cite{Mathijssen2023CulinaryScience}, 
and artificial intelligence \cite{Cichos2020MachineMatter}, 
as discussed in previous review articles \cite{teVrugt2025Metareview:Reviews}.
However, because living systems operate out of equilibrium \cite{Fodor2016HowMatter}, there are still no universal laws that govern how collective phenomena arise, and when they break down.

\begin{figure}[b!]
\centering
\includegraphics[width=\linewidth]{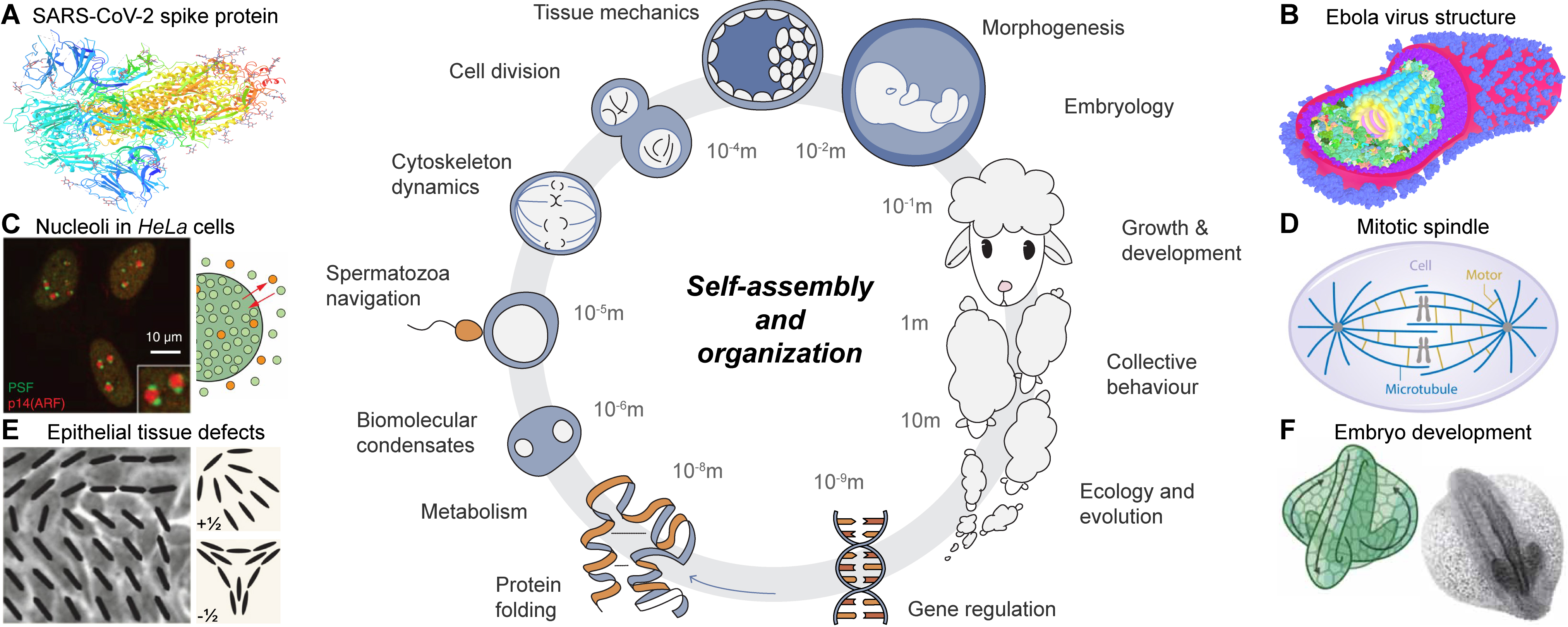}
    \caption{
    \label{fig:SelfOrganization}
        \textbf{Self-assembly, organization, and spatiotemporal coordination.} 
        Artwork by Maggie Liu.
        (\textbf{A}) Structure of the SARS-CoV-2 spike glycoprotein. From NIH 3D (CC-BY).
        (\textbf{B}) Cut-away model of the ebola virus. A helix of protein (yellow) encloses the virus' genetic material (pink). From NIH 3D (CC-BY-NC).
        (\textbf{C}) Liquid-liquid phase separation forming biomolecular condensates, with perinucleolar caps bound to nucleolar bodies. From \cite{Shin2017LiquidDisease}.
        (\textbf{D}) Structure of the mitotic spindle, with polar cytoskeletal filaments actuated by molecular motors, driving chromosome separation. From \cite{Furthauer2022HowMaterials}.
        (\textbf{E}) Defects in the nematic order of epithelial tissues. From \cite{Doostmohammadi2018ActiveNematics}. 
        (\textbf{F}) Tissue morphogenesis in the development of zebrafish embryos. From \cite{Bruckner2024LearningReview}.
    }
\end{figure}

This review focuses on biomedical active matter. 
On the one hand, we discuss how the collapse of collective functionalities can cause diseases and pathological behaviors.
On the other hand, we examine how synthetic active materials can be used to treat health conditions and develop new biomedical devices.
The review is structured in three main chapters, which bridge the scales from molecules to cells [\S\ref{sec:Subcellular}], from cells to tissues [\S\ref{sec:Cellular}], and from tissues to organisms [\S\ref{sec:Organisms}].
In each chapter, we consider four main categories of collective functionalities: 
Self-assembly and spatiotemporal organization [Fig.~\ref{fig:SelfOrganization}];
collective motion, actuation, and transport [Fig.~\ref{fig:Motors}];
collective sensing, signaling, and communication [Fig.~\ref{fig:Sensing}]; and
collective learning, adaptation, and evolution [Fig.~\ref{fig:Learning}].
Inspired by these functionalities, we discuss a wide range of biomedical applications [Fig.~\ref{fig:Biomedical}].
Finally, we highlight the main summary points and future issues in the discussion [\S\ref{sec:Discussion}].

\section{FROM MOLECULES TO CELLS}
\label{sec:Subcellular}

The music of life is conducted by self-regulating biochemical networks that arrange atomic compositions.
In section \S\ref{subsec:SelfAssembly}, we articulate the foundations of how molecular structures acquire function through protein folding and macromolecular engineering. 
In \S\ref{subsec:CellularPartitioning}, we sing of organelles and membraneless biomolecular condensates, which echoes naturally with \S\ref{subsec:GeneRegulation} on chromatin organization and gene regulation. 
In \S\ref{subsec:MolecularMotor}, we tune to cargo transport by molecular motors, resonating with \S\ref{subsec:Organelles} on molecular force generation and \S\ref{subsec:Cytoskeleton} about the dynamics of the cytoskeleton. 
In \S\ref{subsec:IntracellularFlows}, we harmonize the mechanisms of intracellular transport, which sets the scene for \S\ref{subsec:Sensing} on collective sensing and signal transduction. 
Overall, this chapter vocalizes the physical principles that coordinate molecular functions to produce coherent cellular states.
    \subsection{Protein folding and macromolecular engineering}
\label{subsec:SelfAssembly}

Biological systems self-assemble from the nanoscale to meter-sized organisms. At the molecular level, the first key step is protein folding \cite{Dill2012TheOn}, 
where long chains of amino acids turn into functional structures [Fig.~\ref{fig:SelfOrganization}A]. Correctly folded proteins drive essential biological processes, such as DNA replication, signaling, and enzymatic catalysis \cite{Ghosh2021EnzymesMatter}, but misfolding can lead to Parkinson's, Alzheimer's, type-II diabetes mellitus, and prion diseases \cite{Louros2023MechanismsAggregation}.
At the intermolecular level, proteins themselves self-assemble into higher-order structures like protein complexes, fibrils, and viral capsids [Fig.~\ref{fig:SelfOrganization}B]. 

Soft matter and statistical physics can effectively predict protein structures, aggregation, and engineered sequence function \cite{Onuchic2004TheoryFolding}. 
Protein folding typically arises from non‑covalent interactions shaped by temperature, pH, and ionic conditions, captured by funnel‑shaped free energy landscapes that balance entropy and enthalpy. 
Atomistic simulations offer the highest accuracy and agreement with experiments, but they are computationally intensive. 
Coarse‑grained models can be more efficient to explore larger‑scale protein interactions and the general conditions that drive assembly.
Recent AI/ML tools \cite{Cichos2020MachineMatter} like \textit{AlphaFold} and \textit{Rosetta} can also predict protein fold structures from sequences, albeit for limited classes of proteins. 

Today, it is still challenging to model protein folding, especially intrinsically disordered and dynamically moving proteins with kinetic traps.
To capture this, it is essential to consider self-assembly as a non-equilibrium process \cite{Singh2024Non-equilibriumProperties}, opening new opportunities for hierarchically assembled active matter, programmable folding \cite{McMullen2022Self-assemblyFolding}, and self-replication \cite{Hallatschek2023ProliferatingMatter}.
Ultimately, understanding protein folding and self-assembly is a problem at the intersection of physics, chemistry, and biology. Addressing this challenge can lead to new therapeutic strategies for diseases associated with protein misfolding \cite{Louros2023MechanismsAggregation}, and to creating new drugs, vaccines, or antibodies using \textit{de novo} protein design \cite{Watson2023DeRFdiffusion}. 
    
\subsection{Partitioning into organelles and biomolecular condensates}
\label{subsec:CellularPartitioning}

This self-organization continues at the cellular scale through compartmentalization into distinct subdivisions, such as the nucleus and other organelles. 
Partitioning enables cells to control thousands of simultaneous biomolecular reactions with precise spatiotemporal coordination. 
Classical organelles in eukaryotes are membrane bound, but recent developments concern membraneless biomolecular condensates \cite{Aierken2026RoadmapBiology}, regions of highly concentrated macromolecules that remain coherent without separating barrier.
Although the underlying mechanisms are still unclear, these condensates are now thought to be important for physiological processes such as protein aggregation, gene expression, and signal transduction \cite{Banani2017BiomolecularBiochemistry}, while aberrations in partitioning are linked to multiple diseases \cite{Shin2017LiquidDisease}.

The primary driver of biomolecular condensation is liquid-liquid phase separation (LLPS), a demixing process that produces liquid-like droplets of proteins, RNA, or other molecules \cite{Hyman2014Liquid-LiquidBiology}. Due to the lack of a separating membrane, these droplets can rapidly exchange chemicals with their exterior. 
However, this also makes them less stable \cite{Sear2003InstabilitiesComponents}. 
Therefore, it is exciting to explore to what extent LLPS can be triggered, inhibited, or controlled. 
One handle is biological activity, such as localized transcription of RNA, which can induce the formation of nucleoli [Fig.~\ref{fig:SelfOrganization}C]. 
Another handle is actively produced concentration gradients, which generate reaction-diffusion fluxes within the cell.
LLPS may also trigger rheological responses, where parts of the cell become stiffer, altering biological functions \cite{Shin2017LiquidDisease}. 

To advance this area, more work is needed to uncover the mechanisms used by cells to regulate biocondensates, and how this controls functions collectively. Given the high dimensionality of the problem, computer-assisted methods, including Physics-Informed Neural Networks (PINNs) based on Flory-Huggins theory, may prove particularly useful \cite{Cichos2020MachineMatter}.
Another exciting future direction is to control LLPS with active fluids \cite{Tayar2023ControllingFluid}, and to reveal the underlying non-equilibrium dynamics \cite{Aierken2026RoadmapBiology}.

    
\subsection{Chromatin and gene regulation}
\label{subsec:GeneRegulation}

An important example of biomolecular condensation is the organization of the genome into chromatin, complexes of DNA, RNA, and proteins found in the nucleus of eukaryotic cells \cite{Agrawal2017ChromatinMatter, Fierz2019BiophysicsDynamics}. 
Chromatin plays crucial roles in regulating gene expression [Fig.~\ref{fig:Learning}A]:
While prokaryotes predominantly control transcription (from DNA to RNA), eukaryotes also regulate translation (from RNA to proteins). 
Moreover, they use chromatin to control access to genetic information and package DNA into chromosomes [Fig.~\ref{fig:Learning}B]. 
Chromatin exhibits a complex, multiscale architecture that underpins coordinated gene activation and repression \cite{Lawson2023TransposableOrganization, Han2024AdvancesRegulation}, with implications for genome stability, aging, and cancer \cite{Zufferey2021SystematicCancers}. Chromatin is arranged in small units of DNA wrapped around histone proteins, the nucleosomes, which further organize into local domains. At the larger scale, it forms large loops, termed topologically associated domains, that compartmentalize the genome and facilitate interactions between regulatory elements and genes. 
Physically, these functional compartments are, in part, formed through phase separation [\S\ref{subsec:CellularPartitioning}]. Their disruption can lead to genome instability and have dramatic consequences for cell integrity \cite{Lawson2023TransposableOrganization}. For instance, in cancer, defective looping can misregulate key genes, contributing to tumor progression \cite{Zufferey2021SystematicCancers}.

In addition to this structure, chromatin is dynamic and exhibits non-equilibrium fluctuations over a range of temporal and spatial scales. These fluctuations are central to its function, as they provide access to the transcriptional cellular machinery \cite{Fierz2019BiophysicsDynamics}. Both the kinetics and the out-of-equilibrium thermodynamics of chromatin can be successfully described by stochastic active matter and polymer models \cite{Agrawal2017ChromatinMatter, Fierz2019BiophysicsDynamics}. 

A major challenge is determining whether transcriptional changes in cancer cells, such as oncogene activation and tumor suppressor gene silencing \cite{Zufferey2021SystematicCancers}, are a cause or a consequence of the associated chromatin domain alterations. 
More broadly, we still do not know to what extent chromatin folding directly regulates transcription. Some genes within the same domain are strongly co-regulated while others function independently, and loop disruptions do not always alter transcription. This suggests that genome architecture alone does not dictate gene expression \cite{Zufferey2021SystematicCancers}.
Active matter models integrating nonequilibrium physics and stochastic interactions offer promising insights into the interaction between the dynamics and structure of chromatin and gene expression \cite{Agrawal2017ChromatinMatter, Fierz2019BiophysicsDynamics}.

    \begin{figure}[b!]
\centering
\includegraphics[width=\linewidth]{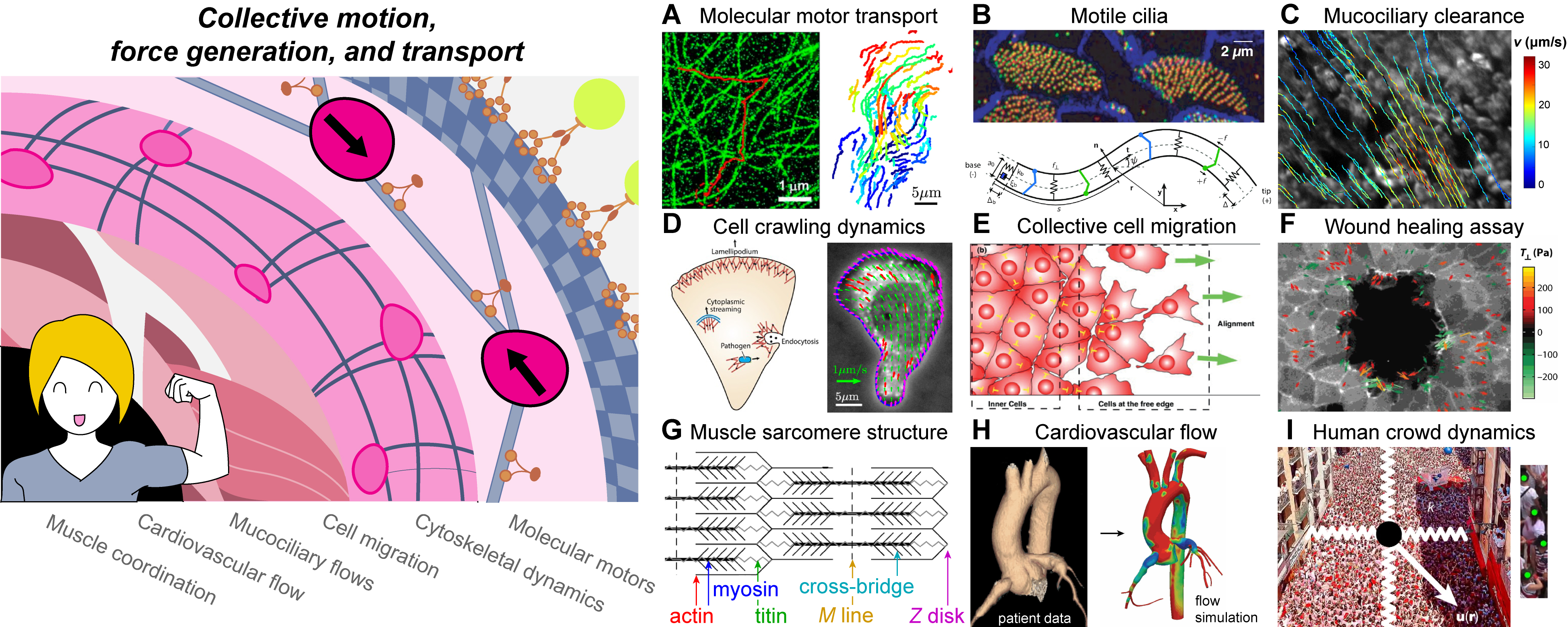}
    \caption{
    \label{fig:Motors}
        \textbf{Collective motion, force generation, and transport.} 
        Artwork by Maggie Liu.
        (\textbf{A}) Cargo transport driven by molecular motors. Left: Lysosome trajectory (red) along microtubules (green) in a monkey kidney cell. Right: Advection of acidified organelles in the flowing cytoplasm of a migrating HL60 cell. From \cite{Mogre2020GettingWorld}.
        (\textbf{B}) Motile cilia. Top: Multiciliated cells in the mouse trachea. From G. Ramirez-San Juan. Bottom: Two-dimensional model of the axoneme. From \cite{Sartori2016DynamicFlagella}.
        (\textbf{C}) Flows driven by cilia in the human airway. From J. Nawroth.
        (\textbf{D}) Cell crawling. Left: Underlying dendritic F-actin networks in the cytoskeleton. From \cite{Banerjee2020TheMaterial}. Right: Flow in migrating neutrophil-like HL60 cell associated with deformation of cell boundary (pink arrows). From \cite{Mogre2020GettingWorld}.
        (\textbf{E}) Collective cell migration, with contact inhibition of locomotion between inner cells (yellow arrows) and polarization of the leaders (green arrows). From \cite{Hakim2017CollectivePerspective}.
        (\textbf{F}) Traction forces in LifeAct-GFP MDCK cells closing a wound. From \cite{Alert2020PhysicalMigration}.
        (\textbf{G}) Schematic of force-generating sarcomeres in muscle myofibrils. Adapted from \cite{Caruel2018PhysicsContraction}.
        (\textbf{H}) Cardiovascular flows from a patient with coronary aneurysms caused by Kawasaki disease. Left: Volume-rendered CT image data. Right: Corresponding flow simulation showing wall shear stress. From \cite{Marsden2014OptimizationModeling}.
        (\textbf{I}) Emergence of collective oscillations in large human crowds. Inset: tracking individual people. From \cite{Gu2025EmergenceCrowds}. 
    }
\end{figure}
    
\subsection{Cargo transport by molecular motors}
\label{subsec:MolecularMotor}

Within cells, cargoes ranging from individual molecules and small vesicles to entire organelles [\S\ref{subsec:CellularPartitioning}] are transported around by molecular motors [Fig.~\ref{fig:Motors}A].
The three classical superfamilies of linear molecular motors are myosin, kinesin, and dynein \cite[see e.g.][]{Mogre2020GettingWorld}. 
Kinesin and dynein travel along microtubules, while myosin travels along actin filaments [see cytoskeleton, \S\ref{subsec:Cytoskeleton}].  
The step size varies greatly among motor species: kinesin and dynein typically take steps of about \SI{8}{\nano\metre}, while some myosin subtypes can stride up to \SI{36}{\nano\metre}. 
Despite these differences, the motors can bind to the same cargo, which facilitates highly non-trivial cooperative effects \cite{Julicher1997ModelingMotors, Chowdhury2006CollectiveCompetition, Holzbaur2010CoordinationDynamics}.

In unidirectional transport, multiple motors may be needed to move large vesicles through the viscous cytoplasm. 
Having multiple motors can prevent detachment events caused by thermal fluctuations \cite{Gross2007CargoOne}. 
The processivity, or average walking distance before detachment, increases as $L_N \approx L_15^{N-1}/N$ for strongly binding motors such as kinesin, where $N$ is the number of bound motors \cite{Klumpp2005CooperativeMotors}. 
This effect allows motors to walk for exponentially longer distances, up to $\SI{300}{\micro\metre}$.

Because dynein and kinesin move in opposite directions, bidirectional transport occurs when these motor families attach to the same cargo. 
Such back-and-forth motion has been observed for organelles, vesicles, and even viruses. 
The most prominent theory to explain this phenomenon is the ``tug-of-war'' model, in which each type of motor competes to control the speed and direction of the cargo \cite{Muller2008Tug-of-warMotors}.
Another proposed explanation is a coordination mechanism, which suggests that only one set of motors is active at a time \cite{Muller2008Tug-of-warMotors}.
Attempts to pinpoint the exact mechanism remain experimentally challenging because of the nature of in vivo experiments, highlighting the complexity of intracellular transport in living systems.






\subsection{Force generation in motile organelles}
\label{subsec:Organelles}

In addition to moving cargo, molecular motors can generate strong forces to power motile organelles, such as cilia and flagella \cite{Gilpin2020TheFlagella}.
Motile cilia are hair-like protrusions from the cell surface [Fig.~\ref{fig:Motors}B] that actively beat to pump fluid flows [Fig.~\ref{fig:Motors}C].
The structure inside these cilia is called the axoneme, where dyneins [\S\ref{subsec:MolecularMotor}] collectively produce shear stresses between microtubule doublets.
These stresses give rise to non-linear oscillations, which in turn lead to asymmetric power and recovery strokes that generate flows at low Reynolds numbers \cite{Bruot2016RealizingColloids}. 
Hence, (multi)ciliated cells in tissues can drive liquid transport, including cerebrospinal fluid circulation in the brain and mucus clearance in the lungs [\S\ref{subsec:CiliaTransport}].

Several interrelated models have been proposed to explain how dyneins are spatially and temporally coordinated, including the geometric clutch model, the sliding control model, and the curvature control model \cite{Gilpin2020TheFlagella, Sartori2016DynamicFlagella}. 
However, it is still unclear how disruptions in dynein coordination are responsible for numerous ciliopathies \cite{Fliegauf2007WhenCiliopathies}. 

These cilia often compete with bacteria, such as \textit{E. coli} trying to invade our lungs by swimming upstream against mucus flows \cite{TorresMaldonado2024EnhancementFluids}.
These bacteria are propelled by helical flagella that do not beat: Instead, they turn like a corkscrew through the fluid, driven by a rotary motor complex \cite{Wadhwa2022BacterialMechanisms}.
This flagellar motor is collectively powered by rotor and stator proteins that convert chemical energy into torques. 
More stators are recruited  when the fluid viscosity increases, so the bacteria can adapt their swimming behavior by using this active matter complex as a mechanical sensor
\cite{Dufrene2020Mechanomicrobiology:Forces}.
Future studies integrating high-speed imaging, force manipulation, protein engineering, and computational modeling will be essential to fully elucidate these motor dynamics and bacterial infection mechanisms.

\subsection{Collective dynamics of the cytoskeleton}
\label{subsec:Cytoskeleton}

The cytoskeleton is a network of actin filaments, microtubules, and intermediate filaments that collectively regulate cellular architecture, movement, mechanical stability, and cell division \cite{Julicher2007ActiveCytoskeleton, Furthauer2022HowMaterials}. 
These fibrous structures provide the foundation for motor proteins [\S\ref{subsec:MolecularMotor}], but the network itself also generates forces by active polymerization and depolymerization. 
Each filament type has distinct physical and biochemical properties: Actin filaments typically generate protrusive and contractile forces, microtubules establish intracellular organization, polarity, and the mitotic spindle [Fig.~\ref{fig:SelfOrganization}D], while intermediate filaments provide mechanical reinforcements and dynamical feedbacks. They physically and biochemically influence each other to coordinate cellular processes in space and time. 

These collective interactions enable cells to crawl [Fig.~\ref{fig:Motors}D] and dynamically reorganize their cytoskeleton in response to external stimuli \cite{Dogterom2019ActinmicrotubuleBiology}.
High-density actomyosin motility assays reveal that filament collisions promote cluster formation, leading to the emergence of large-scale flows that resemble cytoskeletal remodeling during tissue morphogenesis \cite{Friedl2009CollectiveCancer}. Microtubules influence migration plasticity by regulating focal adhesion turnover, a process by which cell adhesion to the extracellular matrix is regulated \cite{Murrell2015ForcingContractility}. Intermediate filaments distribute mechanical stress, allowing cells to maintain cohesion in collective migration  [\S\ref{subsec:CellMigration}]. These interactions demonstrate how cytoskeletal networks balance stability with rapid reorganization, facilitating effective responses to environmental changes that they sense [\S\ref{subsec:Sensing}].

It remains challenging to understand the cytoskeleton, because it is a highly active and adaptive system \cite{Banerjee2020TheMaterial} where biopolymer networks integrate mechanical and biochemical signals to regulate cell behavior. 
These types of cytoskeletal crosstalk have biomedical implications for diseases such as cancer, where metastatic cells exploit cytoskeletal plasticity to transition between migration modes, and in neurodegenerative diseases, where disturbances in the cytoskeleton contribute to cell fragility and degeneration \cite{Dogterom2019ActinmicrotubuleBiology}. Bottom-up active matter approaches emerge as one of the key strategies to control cytoskeletal dynamics and cell migration, to enhance tissue regeneration, and to combat disease progression.



\subsection{Intracellular transport, cytoplasmic streaming, and enhanced diffusion}
\label{subsec:IntracellularFlows}

Molecular motors [\S\ref{subsec:MolecularMotor}], motile organelles [\S\ref{subsec:Organelles}], and the cytoskeleton [\S\ref{subsec:Cytoskeleton}] all interact in a coordinated fashion to enable intracellular transport. 
The collective behavior of these smaller structures can create directed motion that carries large biomolecules and organelles, which is necessary for cell division, shape regulation, and cell migration \cite{Mogre2020GettingWorld, Goldstein2016BatchelorCell, Mogilner2018IntracellularGel}. 

The three main mechanisms of intracellular transport are diffusion, advective flows, and direct transport by linear motors. The latter method allows for fast transport over long distances, such as metre-sized neurons, and the motors can control where specific cargo is sent.
However, it requires more energy input than the other mechanisms.
Diffusion is caused by fluctuations in the liquid, which randomly kick objects around in a random walk. In addition to thermal Brownian motion, these fluctuations can be actively generated as a byproduct of cellular metabolism \cite{Brangwynne2009IntracellularDiffusion}, leading to non-equilibrium diffusion and enhanced molecular fluxes without additional energy expenditure \cite{Guzman-Lastra2021}. However, diffusion is only efficient for moving small objects across comparatively short distances, up to a few hundred microns. 
Long-ranged advective flows can be generated in different ways, such as by cell boundary deformations and the contraction of the cytoskeleton [\S\ref{subsec:Cytoskeleton}].
Moreover, strong flows can be generated by the collective motion of molecular motors that entrain fluid flows in their wake [\S\ref{subsec:Transport}]. 
This is called cytoplasmic streaming, which can carry large cargo over long distances at high speeds \cite{Goldstein2016BatchelorCell}. 
In addition to transport, these flows can carry out different vital functions, such as early spatial organization of oocyte components \cite{Mogre2020GettingWorld}, with precise flow patterning \cite{Dutta2024Self-organizedTwisters}.

Each mechanism still contains many unanswered questions. For motor-driven transport, is cargo sorting organized locally or globally using external signaling? For enhanced diffusion, the degree to which specific cellular activities contribute to energy fluctuations in the cytoplasm remains unknown. For advective flows, how much can cells control flow-patterning, and how important is this mechanism compared to others? Resolving these questions can help treat multiple human diseases \cite{Aridor2000TrafficProcesses}.

\subsection{Sensing and signal transduction in cells}
\label{subsec:Sensing}

\begin{figure}[b!]
\centering
\includegraphics[width=\linewidth]{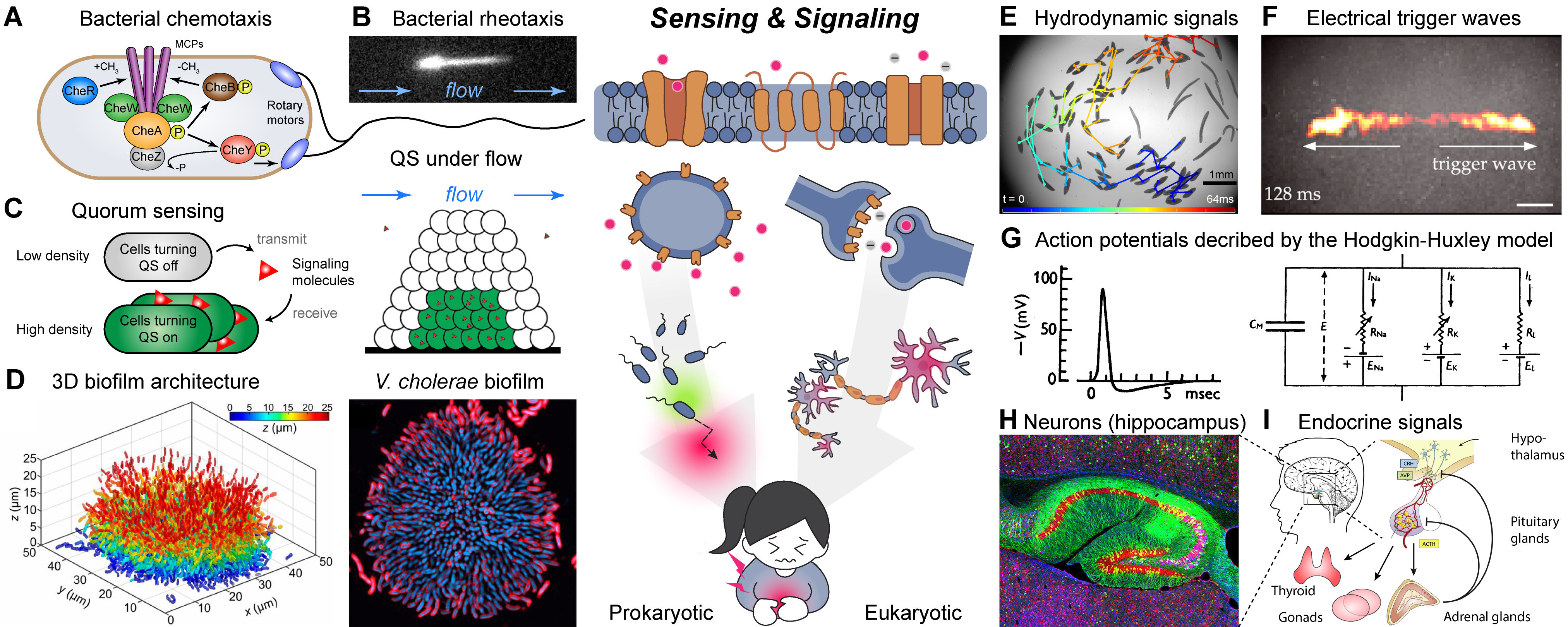}
    \caption{
    \label{fig:Sensing}
        \textbf{Collective sensing, signaling, and communication.} 
        Artwork by Maggie Liu.
        (\textbf{A}) Chemosensory system in \textit{E. coli}. Signals received by MCP receptors are transduced to the flagellar motors.
        (\textbf{B}) Rheotaxis enables bacterial upstream swimming. Courtesy of Ran Tao.
        (\textbf{C}) Bacterial quorum sensing. Right: Signalling under flow. From K. Kim, F. Ingremeau, B. Bassler and H. Stone.
        (\textbf{D}) \textit{V. cholerae} bacterial biofilm architecture. Left: 3D structure imaged with confocal microscopy. Right: Horizontal slice, with nucleoid (blue) and membrane (red) labels. Courtesy of K. Drescher.
        (\textbf{E}) Communication through hydrodynamic trigger waves. From \cite{Mathijssen2019CollectiveWaves}.
        (\textbf{F}) Action potential propagating along tissue interface, revealed by a voltage-sensitive dye. From \cite{Scheibner2024SpikingSystems}.
        (\textbf{G}) Action potentials. Right: Electrical circuit representing a neuron membrane. 
        From \cite{Hodgkin1952ANerve}.
        (\textbf{H}) Mouse hippocampus with neurodegenerative disease Niemann-Pick type C1. From I. Williams, NICHD (CC).
        (\textbf{I}) Hormone signalling pathways in the endocrine system. From OpenStax (CC).
    }
\end{figure}

Cells pose a paradigm for highly efficient sensors [Fig.~\ref{fig:Sensing}], capable of measuring chemical, mechanical, optical, thermodynamic, and other signals. Naturally, their sensing and signal transduction mechanisms are closely intertwined with receptor protein folding [\S\ref{subsec:SelfAssembly}] and intracellular transport [\S\ref{subsec:IntracellularFlows}]. Moreover, instead of using single sensors, cells often use arrays of multiple sensors that collectively detect weak signals with extreme precision, operating close to physical limits \cite{Bialek2005PhysicalSignaling, Mugler2016LimitsIntegration}.

Chemical concentrations and their gradients are processed with subcellular pathways, capable of adaptation and habituation \cite{Eckert2024BiochemicallyLearning} [Fig.~\ref{fig:Learning}C].
Prokaryotes typically use two-component signal transduction systems, composed of histidine kinases and response regulators \cite{Stock2000Two-ComponentTransduction}.
The histidine kinase usually has an extracellular binding domain that detects stimuli (ligands, pH, osmolarity, light) and an intracellular domain that then autophosphorylates.
The phosphoryl group is subsequently transferred to a response regulator that modulates a function, such as redox control, gene expression [\S\ref{subsec:GeneRegulation}], biofilm formation [\S\ref{subsec:Biofilms}], or chemotaxis [\S\ref{subsec:Navigation}].
Eukaryotes also use two-component systems and even more elaborate biochemical networks \cite{Tkacik2025InformationNetworks}.
Ultimately, the physical limit of sensory precision is governed by a signal-to-noise ratio, such as the steepness of the gradient compared to a fluctuating background \cite{Bialek2005PhysicalSignaling}.
To optimize this ratio, cells can compare signals from multiple receptors over time with spatial communication and temporal integration \cite{Mugler2016LimitsIntegration}.
It would be very interesting to implement this in biomimetic microrobots capable of ultra-sensitive detection and navigation [\S\ref{subsec:Navigation}]. 

Mechanotransduction \cite{Cheng2025MechanobiologyTimescales}, when physical stimuli are converted into biochemical signals, provides another key source of information. Mechanosensitive ion channels detect tension in the cell membrane due to deformation. This tension changes the open probability of the channels, leading to an immediate influx of ions. This direct mechanical coupling can lead to extremely fast responses, such as escape behavior, nematocyst firing, and toxin release \cite{Mathijssen2019CollectiveWaves}.  

Single-cell sensing is merely a component of a more complex interaction network. Collective gradient sensing \cite{Camley2018CollectiveDevelopments} and cell-cell communication [\S\ref{subsec:communication}] can improve their detection precision, crucial for collective migration [\S\ref{subsec:CellMigration}]. 
Yet, any cooperative sensing mechanism is still subject to physical limits and information bottlenecks.

    \section{FROM CELLS TO TISSUES}
\label{sec:Cellular}


Drawing upon collective functionalities at the molecular and subcellular scale [\S\ref{sec:Subcellular}], this chapter paints the principles of cooperation between multicellular assemblies and tissues. 
Section \S\ref{subsec:Biofilms} illustrates how bacterial biofilms emerge from communication-driven phenotype switching, while biofilm dispersal prepares the canvas for \S\ref{subsec:Navigation}, which depicts the navigation of cells in complex environments through integrated chemical, mechanical, and hydrodynamic cues. 
These navigation strategies underpin \S\ref{subsec:CellMigration}, which highlights collective cell migration, where coordinated sensing and force generation shape tissue mechanics.
In \S\ref{subsec:MicroroboticSwarms}, we sketch how microrobotic swarms can leverage similar collective behaviors for specific biomedical tasks, such as minimally invasive surgeries and drug delivery. 
\S\ref{subsec:Transport} portrays beneficial corollaries of collective motion, including enhanced mixing and transport by hydrodynamic entrainment.
\S\ref{subsec:CiliaTransport} marks ciliary synchronization and metachronal wave formation, but also their collective sensing abilities.
This leads into \S\ref{subsec:communication}, which outlines cellular communication, information processing, and the development of learning active matter.

\subsection{Bacterial biofilm formation and dispersal}
\label{subsec:Biofilms}

Perhaps the most prototypical example of individual cells working together is the bacterial biofilm \cite{Santos2018WhatPerspective}, which can be found almost everywhere, from primordial microbial mats to pathogens in medical equipment \cite{Mazza2016TheIntroduction}.
Biofilms are dense communities of microorganisms enclosed in a self-produced viscoelastic matrix composed of extracellular polymeric substances (EPS).
This encapsulation offers benefits such as cohesion and adhesion to (sometimes living) surfaces, exclusive access to nutrients and mutual metabolites, and protection against environmental stresses, viruses, and antibiotics \cite{Hall-Stoodley2004BacterialDiseases}.

The early stages of biofilm formation are associated with surface accumulation due to hydrodynamic and steric interactions \cite{Lauga2020TheMotility}, as well as cell-cell interactions and collective motion \cite{Aranson2022BacterialMatter}. 
The next stage involves a phenotype switch from swimming or planktonic cells to sessile communities.
This transition is usually triggered by cell-cell communication [\S\ref{subsec:communication}] through a collective decision making process called quorum sensing \cite{Miller2001QuorumBacteria}. 
To detect whether they have a quorum, bacteria produce and sense their own signaling molecules, called auto-inducers, enabling them to measure their own cell density [Fig.~\ref{fig:Sensing}C].
As they get denser, flagellar motility is downregulated while pili and EPS production are upregulated to promote surface attachment \cite{Wadhwa2022BacterialMechanisms}.
The biofilm then matures, with different bacterial species self-organizing into 3D spatially structured ecosystems [Fig.~\ref{fig:Sensing}D] with rich cooperation and competition dynamics \cite{Nadell2016SpatialBiofilms}. 
In the final stage, bacteria switch back to the planktonic phenotype and disperse, often due to nutrient depletion or other environmental factors \cite{Esteves2025NitricCholerae}.

Studying the mechanisms behind biofilm formation and dispersal can help mitigate infections and contamination of medical equipment. From an active matter point of view, it remains unknown how the biomechanics of collectively moving cells couple to the microrheology of the viscoelastic matrix \cite{Mazza2016TheIntroduction, Gloag2020BiofilmSurvival}. Moreover, while biofilms are often studied in idealized lab conditions, much less is known about their adaptation to dynamical environments, flows, and microconfinement \cite{Conrad2018ConfinedBiofilms}. 
Harnessing this knowledge could enable the design of living materials with tuneable physicochemical functionalities, such as capturing microplastics and other pollutants. 

    
\subsection{Adaptive navigation by cells in complex environments}
\label{subsec:Navigation}


After dispersing from biofilms [\S\ref{subsec:Biofilms}], bacteria and other microswimmers \cite{Elgeti2015PhysicsReview} can actively navigate through complex environments using ``tactic'' responses, including chemotaxis, phototaxis, viscotaxis, durotaxis, and rheotaxis \cite{Ishikawa2025PhysicsStimuli}.
These steering mechanisms are internally regulated, allowing cells to adapt their responses.

Chemotaxis, for example, allows cells to move towards nutrient sources or move away from toxins \cite{Wadhams2004MakingChemotaxis}.
\textit{E. coli} bacteria use a two-component signal transduction system [\S\ref{subsec:Sensing}] with one histidine kinase, CheA, a transmembrane receptor that activates two response regulators [Fig.~\ref{fig:Sensing}A]: 
CheY controls the flagellar motors, which leads to run-tumble motion \cite{Wadhwa2022BacterialMechanisms}.
CheB regulates receptor demethylation, allowing the cell to reset its sensitivity to chemical gradients adaptively \cite{Tu2013QuantitativeAdaptation}.
Interestingly, using arrays of chemoreceptors collectively, bacteria can detect changes of just a few molecules over background concentrations that can vary over five orders of magnitude \cite{Wadhams2004MakingChemotaxis}.
This is extremely important for microbial ecology in nutrient-poor environments \cite{Keegstra2022TheChemotaxis}.
To achieve this, they use fold‑change detection or logarithmic sensing: 
The cell response depends on $\Delta(\log c)=\Delta c/c$, rather than $\Delta c$. 
The same idea holds for phototaxis, where the sensitivity also scales with relative changes in stimulus intensity, not absolute changes. 
This logarithmic sensing is analogous to the Weber–Fechner law in human perception, including vision, hearing, smell, taste, and touch.

Bacteria can also respond to mechanical and hydrodynamic cues \cite{Persat2015TheBacteria}.
Rheotaxis is the directed movement of cells in fluid currents \cite{Wheeler2019NotPlankton}:
Instead of going with the flow, bacteria can swim upstream [Fig.~\ref{fig:Sensing}B] to invade anatomical tracts and biomedical devices \cite{Tao2026InvasionDevices}. 
Viscotaxis is the ability to move to regions of higher or lower viscosity, while durotaxis allows cells to climb stiffness gradients.
Moreover, the bacterial motors act as mechanosensors with force-induced feedback on motility and adhesion \cite{Dufrene2020Mechanomicrobiology:Forces}.
Together, these mechanisms are essential for motility in complex physiological environments, such as mucus in the gut or lungs. 

Indeed, multisensory perception is critical for navigation in any realistic habitat. 
However, it remains unclear how these tactic responses interact with each other. 
For example, chemotaxis and rheotaxis are coupled because of ligand advection in flows. 
Therefore, they can enhance each other if the flow and chemical gradients are aligned, which can help bacteria find distant upstream nutrient sources.

\subsection{Collective cell migration and active tissue mechanics}
\label{subsec:CellMigration}

While individual microswimmers are excellent navigators  [\S\ref{subsec:Navigation}], they often cooperate to detect signals better as groups \cite{Camley2018CollectiveDevelopments}. 
Collective cell migration [Fig.~\ref{fig:Motors}E] underpins numerous processes ranging from tissue mechanics and embryo morphogenesis [Fig.~\ref{fig:SelfOrganization}E,F] to the spread of cancer \cite{Friedl2009CollectiveCancer}.  
In order to reliably control these forms of collective behavior, it is vital to understand the physical and biological mechanisms behind cell migration, for example to inhibit metastasis or promote wound healing [Fig.~\ref{fig:Motors}F].

Collective cell migration can take the form of moving monolayers, loosely connected liquid-like streams, small clusters, or complex 3D structures \cite{Alert2020PhysicalMigration}. 
Cell movement is guided by environmental cues and constraints, which can take the form of spatial confinement, substrate stiffness, tissue topography, and rheological properties \cite{Hakim2017CollectivePerspective}. 
Motility is often driven by the formation of plasma membrane protrusions, such as filopodia and lamellipodia. These organelles can be formed by actin polymerization or intracellular pressure gradients, and they anchor the cells to their environment, providing the traction necessary for locomotion \cite{Alonso-Matilla2025PhysicalMigration}.

Despite growing interest in active tissue mechanics \cite{Xi2018MaterialMechanics}, many questions remain: 
At smaller scales, what molecular interactions are responsible for the forces between cells and the substrate? 
At larger scales, how do external cues such as electrical, chemical, and mechanical signals affect the mechanisms of migration? 
To approach these questions, theoretical models can be categorized as bottom-up and top-down theories \cite{Bruckner2024LearningReview}:
Bottom-up models aim to integrate the key mechanisms at play to render quantitative predictions. Examples include phase field, agent-based, molecular clutch, cellular Potts, active gel, vertex, and active nematics models \cite{Doostmohammadi2018ActiveNematics}. 
Top-down models, conversely, infer predictions from experimental data agnostic to microscopic details. Examples include simple regression models, Bayesian approaches, and large spatiotemporal AI and self-learning models analogous to \textit{AlphaFold} [\S\ref{subsec:SelfAssembly}].
For these models to be successful, they will need to be coupled with much more \textit{in vivo} and \textit{in vitro} experiments, requiring the development of more advanced imaging and analysis techniques.
    
\subsection{Microrobotic swarms for drug delivery and minimally invasive surgeries}
\label{subsec:MicroroboticSwarms}

Collective motion [\S\ref{subsec:CellMigration}] can also be exploited in synthetic systems, such as microrobotic swarms. 
These swarms offer significant advantages in biomedical applications \cite{Ghosh2020ActiveTherapeutics}, particularly in drug delivery and minimally invasive surgeries \cite{Nelson2010MicrorobotsMedicine, Sitti2015BiomedicalMilli/Microrobots}. 

Unlike individual microrobots, swarms exhibit emergent behaviors \cite{Zottl2016EmergentColloids} such as load sharing, specialization, and adaptive responsiveness, enabling efficient task execution in physiological environments \cite{Yang2022AMicrorobotics}. 
In drug delivery, they navigate complex vascular networks [\S\ref{subsec:cardiovascular}], collectively overcome obstacles, and accurately release drugs to reduce side effects [Fig.~\ref{fig:Biomedical}C,D]. 
In micro-surgeries, they enable embolization, tissue removal, targeted cauterization with reduced unwanted damage, and biofilm removal [Fig.~\ref{fig:Biomedical}H]. 
%
Recent advances in biohybrid microrobots integrate synthetic materials with biological components, broadening their applications and increasing biocompatibility. 
Furthermore, active particles can be guided by endogenous fields, such as chemical gradients produced by tumors \cite{Lin2025InteractionsFields}.
In addition, real-time imaging modalities such as MRI and ultrasound facilitate precise in vivo monitoring.
Not least, autonomous coordination and navigation can be improved by collective swarm intelligence \cite{Bonabeau1999SwarmIsystems}.

However, these collective functionalities often break down in realistic environments.
To optimise the design of microrobotic swarms in physiologically relevant settings, it is essential to understand their microscopic dynamics \cite{Bechinger2016ActiveEnvironments}. 
Specifically, we must develop effective theoretical models that capture robot-robot and robot-environment interactions, especially in complex microconfinement and fluid flows \cite{Lauga2020TheMotility}.
Indeed, studies on the physics of microswimmers \cite{Elgeti2015PhysicsReview}, motile active matter \cite{Gompper2025TheRoadmap}, and the biology of microbial navigation [\S\ref{subsec:Navigation}] could provide critical insights.
Future research should also focus on integrating bio-inspired coordination strategies with advanced control algorithms, real-time sensing, and adaptive behaviors to enhance the reliability and efficacy of microrobotic systems.

\subsection{Hydrodynamic entrainment and mixing due to collective motion}
\label{subsec:Transport}

Directed and collective motion [\S\ref{subsec:Navigation}-\ref{subsec:MicroroboticSwarms}] can give rise to the transport of non-motile particles: 
Due to viscous drag, moving objects entrain the fluid and particles around them, like tree leaves in the wake of a car. 
This entrainment is called Darwin drift, after Sir Charles G. Darwin, the grandson of the famous biologist.

Because of this effect, collectively moving microrobots or cells can generate long-ranged fluid flows \cite{Jin2021CollectiveMicroswimmers}. 
For example, collectively moving molecular motors drive cytoplasmic streaming  [\S\ref{subsec:IntracellularFlows}] and microbial active carpets can produce nutrient currents  \cite{Mathijssen2018NutrientCarpets}. 
Moreover, hydrodynamic entrainment can lead to large-scale mixing, the prevention of sedimentation, and enhanced diffusion with L\'{e}vy flights  \cite{Kanazawa2020LoopySuspensions}. 

This transport by active populations can play an important role in microscopic and macroscopic ecosystems. 
For example, viruses and other non-motile pathogens could use this effect to hitchhike and enhance their dispersal.
In oceans, zooplankton and fish have both been proposed as important sources of oceanic mixing, in particular through diel vertical migration \cite{Dabiri2024DoOcean}. However, further experimental and theoretical work is needed to quantify the relative role of biogenic mixing. 
Not least, enhanced transport and mixing can strengthen cooperative interspecies interactions, for instance, between fungi and motile bacteria \cite{Aranson2022BacterialMatter, Keegstra2022TheChemotaxis}.

These interactions between active matter and its environment have also inspired applications. Notably, using motile microorganisms has been proposed to enhance inoculant mixing in agriculture for enhanced crop yield.
Moreover, active transport could be used for drug delivery, where collective entrainment could amplify the effective cargo capacity over an order of magnitude \cite{Jin2021CollectiveMicroswimmers}.






\subsection{Metachronal wave transport and sensing by cilia}
\label{subsec:CiliaTransport}

Motile cilia are another efficient mechanism for active transport \cite{Bruot2016RealizingColloids}.
They are found almost everywhere the human body, including the airways, fallopian tubes and brain ventricles, where they perform essential biological functions such as fluid transport, mucociliary clearance, sperm motility, and breaking the left-right symmetry in developing embryos \cite{Gilpin2020TheFlagella}. 


Analogous to molecular motors collectively driving a single cilium [\S\ref{subsec:Organelles}], the cilia themselves also work together in unison:
They synchronize their beating and form metachronal waves \cite{Wan2024MechanismsCoordination}.
The mechanisms behind this self-organization span across multiple lenghtscales.
At the subcellular level, the cilia are physically coupled to each other through basal bodies and cytoskeletal linkages. 
At the cell scale, metachrony is driven by steric and hydrodynamic interactions between neighbouring cilia.
At multicellular scale, ciliary coordination can be regulated further by paracrine signals, electrical membrane potentials, and neuronal stimuli \cite{Marinkovic2020NeuronalFeeding}.
Therefore, ciliated organisms can rapidly switch their metachronal wave patterns and change their swimming direction within milliseconds \cite{Kourkoulou2025MetachronalPredators}. 

Additionally, non-motile or primary cilia act as precise biological sensors \cite{Spasic2017PrimaryFunction}.
Like antennas, they are loaded with biochemical receptors [\S\ref{subsec:Sensing}], and they can detect mechanical signals such as tiny tissue deformations and extracellular fluid flows:
When the primary cilia bend, they use mechanosensitive ion channels to initiate mechanotransduction cascades that convert physical cues to electrochemical signals and neuronal feedback [\S\ref{subsec:transduction}]. 

However, ciliary dysfunction can lead to many different ciliopathies \cite{Fliegauf2007WhenCiliopathies}, including neurodevelopmental disorders, infertility, and airway diseases, such as cystic fibrosis (CF) and primary ciliary dyskinesia (PCD).
Underlying problems can be in the axoneme structure and dynein motor coordination, basal body or cytoskeletal defects, electrophysiological function loss, or abnormal mucus rheology.
Unravelling these pathological mechanisms can facilitate new therapeutic strategies and diagnostics.
Moreover, inspired by the versatility of cilia, there is an explosion of research in the field of artificial cilia [\S\ref{subsec:BioinspiredMaterials}]. 

    \begin{figure}[b!]
\centering
\includegraphics[width=\linewidth]{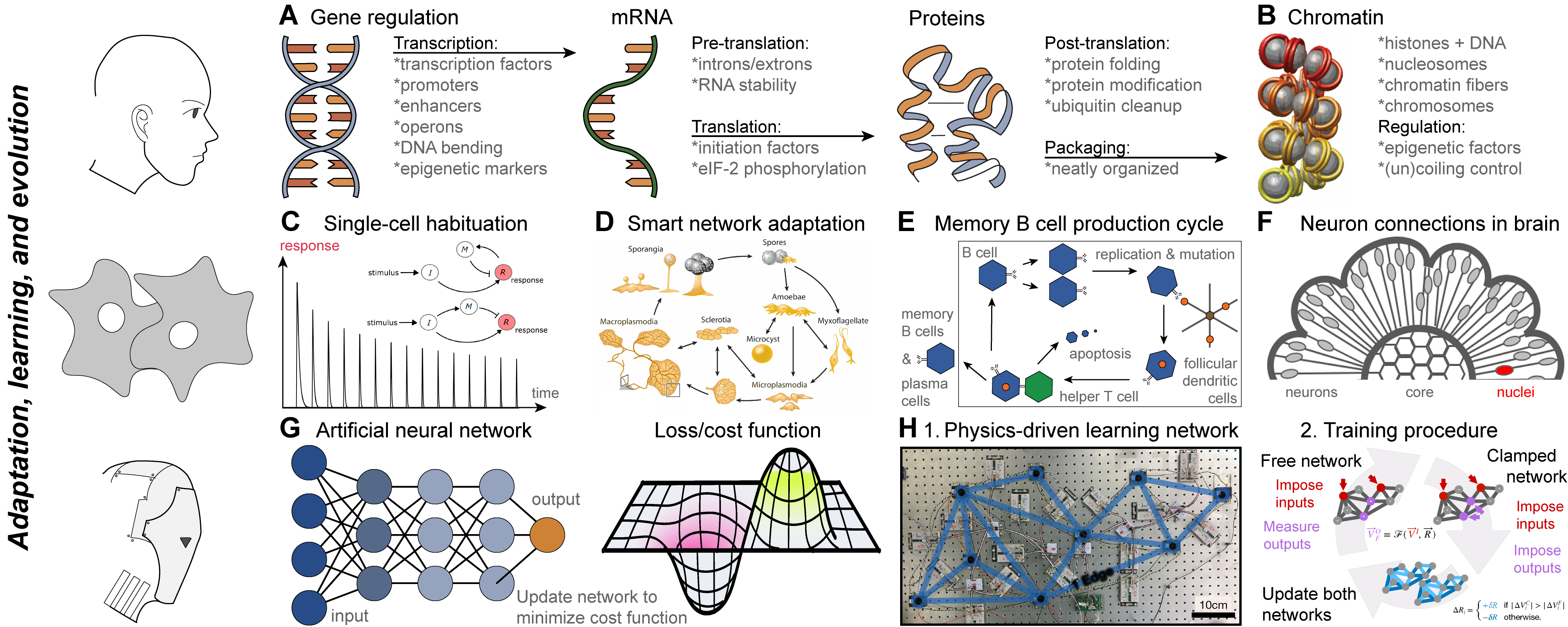}
    \caption{
    \label{fig:Learning}
        \textbf{Collective learning, adaptation, and evolution.} 
        Artwork by Maggie Liu.
        (\textbf{A}) Overview of mechanisms that regulate gene expression. 
        (\textbf{B}) Chromatin gene regulation, by packaging DNA into tight coils around histone proteins. Image from \cite{Fierz2019BiophysicsDynamics}.
        (\textbf{C}) Single-cell learning by habituation to repetitive stimuli. Insets: Biochemical networks that implement habituation. From \cite{Eckert2024BiochemicallyLearning}.
        (\textbf{D}) Smart network adaptation in the giant slime mold \textit{Physarum polycephalum}. From \cite{LeVerge-Serandour2024PhysarumAdaptation}.
        (\textbf{E}) Immune system adaptation: the affinity maturation cycle where memory B cells proliferate and develop receptor mutations, which is evaluated using helper T cells. Adapted from \cite{Chakraborty2017RationalMedicine}.
        (\textbf{F}) Neurons forming connections in the brain. From Eyal Karzbrun and Orly Reiner.
        (\textbf{G}) Artificial neural networks, trained by minimizing a cost function using algorithms such as backpropagation. 
        (\textbf{H}) Physics-driven learning machine: Left: Electical circuit with 16 adaptive resistors. Right: This network is trained to achieve desired output voltages for a given input. Adapted from \cite{Dillavou2022DemonstrationLearning}. 
    }
\end{figure}
    
\subsection{Communicating and learning active matter}
\label{subsec:communication}

Most of the collective functionalities described above can only be achieved through active communication \cite{Ziepke2022Multi-scaleMatter}. 
The exchange of information between cells, tissues, and organism populations establishes a foundation for social interactions \cite{Song2019CellcellOpportunity}. 

Cell-cell communication is initiated by signal transduction at the subcellular level [\S\ref{subsec:Sensing}], which subsequently enables interactions within and between species, including bacterial quorum sensing [\S\ref{subsec:Biofilms}].
These interactions then drive the emergence of cooperative responses, including collective sensing, bioluminescence, biofilm formation, virulence factor production, and the release of public goods that are shared by the community \cite{Bridges2022SignalBehaviors}. %

Akin to biochemical networks enabling the adaptation of individual cells [\S\ref{subsec:Navigation}], networks of communicating cells can cooperatively adapt and ``learn'' [Fig.~\ref{fig:Learning}D].
For example, the social amoeba \textit{Dictyostelium} uses chemical communication to regroup and adapt to brutal environmental changes, and the giant slime mold \textit{Physarum} forms smart adaptation networks \cite{LeVerge-Serandour2024PhysarumAdaptation}. 
Similarly, the human immune system remembers pathogens previously encountered using a memory B cell production cycle [Fig.~\ref{fig:Learning}E].
Understanding these learning cycles could allow us to develop better vaccine targets \cite{Chakraborty2017RationalMedicine}.  

The speed of biochemical communication is intrinsically limited to molecular diffusion. Therefore, cells have also evolved ways to communicate using mechanical, electrical, and optical signals.
The protist \textit{Spirostomum} uses mechano-sensitive ion channels to transmit long-ranged hydrodynamic trigger waves from cell to cell [Fig.~\ref{fig:Sensing}E].
Likewise, neurons use voltage-dependent ion channels to transmit action potentials as electrical trigger waves [Fig.~\ref{fig:Sensing}F], as described by the famous Hodgkin-Huxley model [Fig.~\ref{fig:Sensing}G].
A key difference is that the latter has synapses with inhibitory and excitatory neurotransmitters to regulate complex feedback loops.
This, in turn, facilitates the emergence of coherent states of collective neuronal firing patterns \cite{Eckmann2007TheNetworks}. 
Understanding this spontaneous activity of neuronal networks is crucial for neuroscience and whole-body information processing [\S\ref{subsec:transduction}], and it has inspired a new field of physics-driven learning networks \cite{Stern2023LearningSystems} [Fig.~\ref{fig:Learning}H].


    
\section{FROM TISSUES TO ORGANISMS}
\label{sec:Organisms}


Having analyzed collective behaviors within tissues [\S\ref{sec:Cellular}], this chapter investigates the foundations of coordination at the organismic scale. 
In \S\ref{subsec:cardiovascular}, we explore cardiovascular and other flow networks, mapping out how self-organized architectures and adaptive vessel dynamics efficiently distribute nutrients and regulate whole‑body transport. 
\S\ref{subsec:transduction} examines information transduction between tissues and organs, inquiring how endocrine, neural, and immune signaling interact through nonlinear feedback and critical behavior to maintain homeostasis.
\S\ref{subsec:muscleCoordination} probes how muscular coordination emerges from multiscale feedback, and \S\ref{subsec:Organismic} extends this to multi-organism collective motion and epidemiology. 
Moving toward technological applications, \S\ref{subsec:BioinspiredMaterials} and \S\ref{subsec:Metafluids} scrutinize biomimetic materials, metamaterials, and metafluids inspired by these biological systems. 
Finally, \S\ref{subsec:diagnosticDevices} and \S\ref{subsec:Organoids} delve into cooperative biosensors and organoids-on-chips, which leverage active and self-organizing components to improve sensing, model disease, and enable patient-specific medicine.

\subsection{Cardiovascular transport and other flow networks}
\label{subsec:cardiovascular}

Living systems take advantage of flow networks to distribute material over long distances. 
Examples include plant xylem and phloem, fungal mycelia, and the human lymphatic and cardiovascular systems \cite{Marsden2014OptimizationModeling} [Fig.~\ref{fig:Motors}H].
In each of these, the flow network must effectively service the entire body with minimal material and energy cost \cite{Ronellenfitsch2016GlobalNetworks}.
Therefore, universal scaling laws arise that govern the structure, size, and behavior of vascular networks. 
In mammals, metabolic rate scales with body mass as a power law with exponent 3/4, which can be understood as a trade-off between the material cost for vessel walls and viscous dissipation. 

Moreover, these flow networks must self-organize because the exact location of smaller vessels cannot be efficiently encoded genetically.
Simple design schemes such as Murray's law can recover both hierarchical and reticulated phenotypes, indicating that universal classes of network structures may arise from self-organization \cite{Ronellenfitsch2016GlobalNetworks}.
Yet, because the heart pumps in alternating cycles, systole and diastole, blood flow is pulsatile in the arteries. 
This changes the calculations of power consumption, especially with elastic vessel walls and complex blood rheology \cite{Secomb2016Hemodynamics}.

To regulate flows, biological networks are often capable of modulate their local vein diameters.
For example, \textit{Physarum} can be modeled as a smart adaptive network \cite{LeVerge-Serandour2024PhysarumAdaptation}, where balancing hydrodynamic shear stresses with actomyosin‑generated active stresses predicts which veins grow or shrink. 
Such adaptive network models are crucial to understand vascular transport in health and disease, but they also provide new opportunities for self-regulating biomedical devices [\S\ref{subsec:diagnosticDevices}] and organoids [\S\ref{subsec:diagnosticDevices}].


\subsection{Information transduction between tissues and organs}
\label{subsec:transduction}

Multicellular organisms take advantage of these flow networks [\S\ref{subsec:cardiovascular}] to send robust signals throughout the organism. The endocrine system [Fig.~\ref{fig:Sensing}I] provides a vast array of hormone circulation, balanced by nonlinear feedback loops that can respond over a wide variety of timescales and crosstalk interactions that generate oscillations, bistability, and pathological transitions. Information is encoded not only in hormone concentration but also in pulse frequency and amplitude, and disease can arise from critical shifts in the feedback loop rather than gradual component failure \cite{Zavala2019MathematicalSystems}.

In parallel, organisms also use electrical signals to transmit information between tissues and organs [Figs.~\ref{fig:Sensing}H, \ref{fig:Learning}F]. These electrical signals are typically carried in neurons, which facilitate fast communication across long distances [\S\ref{subsec:communication}]. Neural culture experiments have provided means of evaluating transmission reliability and demonstrated that rate-coded signals decay primarily through accumulated noise along the network. Crucially, stable long-ranged signaling depends on an excitatory/inhibitory balance: Disabling inhibition reduces information transmission despite increased neuron activities, indicating that organism-scale neural signaling operates near self-organized critical gain conditions \cite{Eckmann2007TheNetworks}. Such adaptive coupling also motivates large scale physical learning systems [Fig.~\ref{fig:Learning}H], in which physical substrates adapt internal parameters through local rules to perform computation and memory without centralized control \cite{Stern2023LearningSystems}.

The immune system forms a third pillar for information transduction, which combines long range cytokine signaling with mobile cellular agents. Statistical physics often treats immunity as an adaptive control system \cite{Perelson1997ImmunologyPhysicists}. Clonal expansion provides nonlinear amplification, while antigen depletion and network interactions provide feedback stabilization [Fig.~\ref{fig:Learning}E]. Repertoire recognition can be framed geometrically in shape space, linking microscopic binding to macroscopic response. Within tissues, immune information spreads in time and space. Interestingly, the spatial distributions of immune cells in tumors other than merely their density are correlated with clinical outcomes, suggesting that geometry itself carries functional information \cite{Yu2021PhysicsTumors}.







    
\subsection{Coordination of muscles and biolocomotion}
\label{subsec:muscleCoordination}

These information transduction mechanisms control our everyday movements, through the coordination of muscles [Fig.~\ref{fig:Motors}G]. 
At the microscopic scale, muscles are composed of individual fibers that are arranged into bundles of fascicles that collectively form the muscle belly. 
These fibers contain the actomyosin machinery that powers contraction [\S \ref{subsec:MolecularMotor},\ref{subsec:Cytoskeleton}]. Whereas a single myosin motor only walks a few microns per second, a human arm can move a million times faster by telescopic amplification. Together, these combined forces allow our muscles to serve as the body's motors, brakes, springs, and struts \cite{Caruel2018PhysicsContraction}.  

In human biolocomotion, the macroscopic coordination of the musculoskeletal system enables a wide variety of motions and gaits \cite{Zajac2003BiomechanicsWalking}. A person standing on both legs involves coordinated torques generated by the left and right hip abductors and adductors, with antagonistic control. 
The mechanical linkage of musculoskeletal structures also causes collective motions of the body, or coactivation. 
For example, human knee flexion results in simultaneous hip joint flexion. 
Every motion requires specific muscle coordination patterns. 
Importantly, optimal muscle coordination can switch between minimal energy expenditure and maximal power output with collective load sharing \cite{Caruel2018PhysicsContraction}. 

From a clinical point of view, there are exciting new developments in the field of biomechanics to enhance muscle coordination in patients with muscle diseases, gait impairments, and amputees \cite{Zajac2003BiomechanicsWalking}.
Moreover, the mean life expectancy for patients with Duchenne muscular dystrophy has increased around ten years compared to two decades ago.
However, it remains poorly understood how the central nervous system regulates muscle coactivation.
Biorobots are now being used to emulate and investigate the neuromechanics of animal locomotion \cite{Ramdya2023TheBack}, leading to more effective orthopaedic surgical procedures, wearable robotics, pacemakers, and the use of active materials in prosthetics and orthotics  \cite{Tucker2015ControlReview}.

\subsection{Organismic collective behaviour and epidemiology}
\label{subsec:Organismic}

In addition to coordinated locomotion within a single organism [\S\ref{subsec:muscleCoordination}], we have all seen collective motion in bird flocks, fish schools, and other groups of organisms.
These grand displays naturally emerge from simple local interactions \cite{Vicsek1995NovelParticles, Toner1995Long-RangeTogether}.
This facilitates collective animal behaviors, social dynamics, and complex biological functions that immediately impact fitness and survival \cite{Ouellette2022ABehavior}.
Of all species, humans have developed the most advanced forms of self-organization, with immense benefits.

However, there are important health risks associated with collective behavior:
The emergence of collective oscillations in massive human crowds [Fig.~\ref{fig:Motors}I] can lead to people being physically compressed, as seen in the 2010 Love Parade disaster \cite{Gu2025EmergenceCrowds}.
Group behavior driven by social norms can also drive waves of infection, for example, if people collectively decide not to use vaccines \cite{Morsky2023TheSpread}.
Additionally, social interactions between individuals impacts how a group collectively moves \cite{Ling2019CostsFlocks}.

\begin{figure}[b!]
\centering
\includegraphics[width=\linewidth]{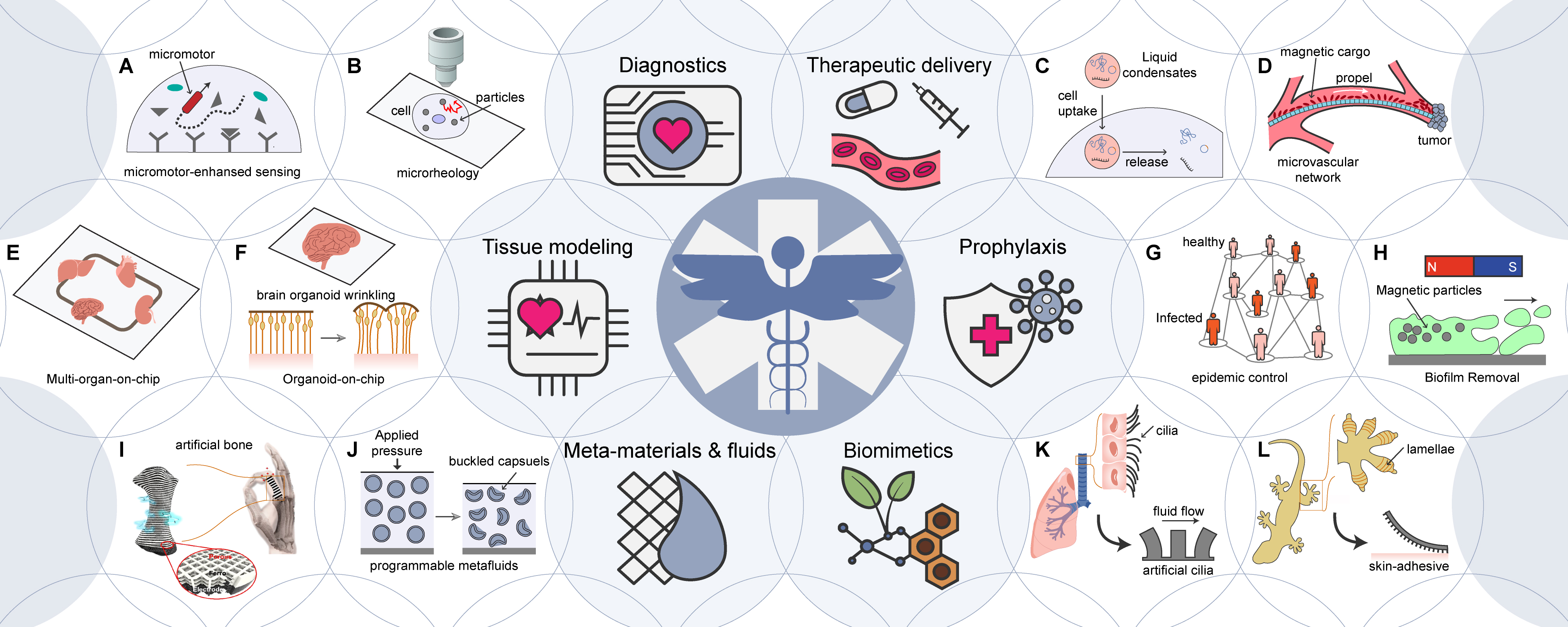}
    \caption{
    \label{fig:Biomedical}
        \textbf {Biomedical applications of collective functionalities}. 
        Artwork by Hamed Almohammadi and Maggie Liu.
        Examples include (i) Diagnostics, with 
        (\textbf{A}) active particles enhance biosensing and 
        (\textbf{B}) single-cell microrheology to detect pathological conditions; 
        (ii) Therapeutic delivery, including 
        (\textbf{C}) synthetic peptide condensates for intracellular cargo delivery and 
        (\textbf{D}) artificial microtubules, adapted from \cite{Gu2022ArtificialMicrocargoes}; 
        (iii) Tissue modeling, comprising 
        (\textbf{E}) multi-organ-on-a-chip systems, and 
        (\textbf{F}) brain organoids, adapted from \cite{Karzbrun2018HumanFolding};
        (iv) Prophylaxis, involving 
        (\textbf{G}) collective disease spread prevention and 
        (\textbf{H}) biofilm removal with magnetic nanoparticles; 
        (v) Metamaterials and metafluids, including 
        (\textbf{I}) 3D-printed piezoelectric artificial bones, adapted from \cite{Li2021BulkManufacturing}, and
        (\textbf{J}) fluids with programmable optical and mechanical properties, adapted from \cite{Djellouli2024ShellMetafluids}; 
        and (vi) Biomimetics, such as 
        (\textbf{K}) artificial cilia for microfluidic flow manipulation and 
        (\textbf{L}) gecko-inspired adhesive patches.
    }
\end{figure}

Moreover, collectively moving pathogens or their vectors, such as swarming bacteria, migrating mosquitoes, or traveling humans, can also enhance disease transmission [Fig.~\ref{fig:Biomedical}G].
Indeed, human collective motion has implications in epidemiology.
A model of infection spread based on WiFi data from crowded events reveals that intermittent human motion patterns, between walking and standing still, create heavy-tailed contact duration distributions, which in turn can increase infection probabilities \cite{Rutten2022ModellingEvents}.

Together, these works reveal interesting connections between social dynamics, tipping points in epidemiology, and implications for public health policy. Future research could develop new protocols to avoid crowd quakes, inform infection control measures, and prevent points of no-return.







\subsection{Biomimetics and bioinspired materials}
\label{subsec:BioinspiredMaterials}

Nature has inspired a vast range of biomimetic design strategies for biomedical applications \cite{Wei2024BiomimeticApplications}, ranging from photonic crystals and sharp structures to self-cleaning surfaces and biocompatible adhesives [Fig.~\ref{fig:Biomedical}L].
While the majority of these are static, a small subset has dynamic properties, and even fewer are active materials with collective functionalities.

At the nanoscale, biomolecular machinery can be reconstituted to produce programmable materials. 
For example, contractile proteins can be used to build ultra-fast and light-controlled chemomechanical networks, with applications in microparticle transport and artificial muscles \cite{Lei2026Light-inducedNetworks}.
Similarly, protein engineering can be used to modify molecular motors or develop \textit{de novo} proteins with previously unexplored functions \cite{Watson2023DeRFdiffusion}.

At the microscale, we already discussed robotic swarms [\S\ref{subsec:MicroroboticSwarms}] and drug delivery by hydrodynamic entrainment [\S\ref{subsec:Transport}]. 
Another class of systems is that of physics-driven learning machines \cite{Dillavou2022DemonstrationLearning}.
Examples include self-regulating microfluidic, mechanical, or electrical networks that are capable of allosteric communication, resource distribution, and dynamic adaptation.
Unlike neural networks \cite{Cichos2020MachineMatter} that require a central processor to minimize a global cost function [Fig.~\ref{fig:Learning}G], these decentralized systems learn by locally tuning interactions between network nodes [Fig.~\ref{fig:Learning}H].

At the millimetre scale, there is much interest in active materials that mimic cilia [\S\ref{subsec:CiliaTransport}].
These artificial cilia [Fig.~\ref{fig:Biomedical}K] can be driven magnetically, pneumatically, acoustically, optically, or electrochemically \cite{Wang2022CiliaManipulation}, with applications in microfluidic mixing, particle transport, self-cleaning surfaces, and microrobot locomotion \cite{ulIslam2022MicroscopicReview}.
Fabrication methods range from nanolithography to 3D printing, and advanced techniques are also capable of producing metachronal waves \cite{Cui2024MetachronalCilia}. 
However, because most artificial cilia are driven by global fields, an important shortcoming is the lack of local control.
Moreover, it remains extremely challenging to truly mimic biological cilia because of their small size and kinematic complexity \cite{Peerlinck2023ArtificialNature}.
``There is plenty of room at the bottom.''

At the centimeter scale, actuatable hydrogels are used for regenerative medicine \cite{Tong2021AdaptableCells}. 
These adaptive materials can be attached directly to tissues to provide a dynamic mechanical microenvironment.
This can influence specific cell behaviors, including cell growth, differentiation, expansion, and migration; with applications in wound healing, inflammation inhibition, spinal cord recovery, and the regeneration of bone, cardiac, cartilage, and osteochondral tissues. 
Besides natural tissues, this technology can be applied to synthetic tissue engineering and organoids [\S\ref{subsec:Organoids}].

\subsection{Active metamaterials and metafluids for biomedicine}
\label{subsec:Metafluids}

Another strategy for developing biomedical applications is to use metamaterials, which exhibit functions that naturally arise from their structure \cite{Bertoldi2017FlexibleMetamaterials}. 
This can result in surprising properties, such as a negative Poisson's ratio, which has also been observed in living materials such as cat skin and cancellous bone.
These properties present exciting opportunities as fundamentally new building blocks in biotechnology \cite{Wang2023AEngineering}.
For example, metamaterial bone implants [Fig.~\ref{fig:Biomedical}I] use auxetic microstructures to mimic natural bone properties, integrate conductive layers to harvest energy, perform sensing, and adapt to accommodate patient needs throughout the healing process. 
Moreover, smart bandages have been proposed as a form of adaptive auxetic metamaterial, which can sense swelling and respond by releasing wound-healing agents. 
 
Similarly, metafluids have recently come into focus as a new avenue for manipulating structure-function relationships in fluidic systems [Fig.~\ref{fig:Biomedical}J].
Unlike conventional fluids, these systems often have remarkable properties.
For example, acoustically driven dense colloidal suspensions exhibit a programmable viscosity, 
and buckling shells feature a programmable compressibility and optical behavior \cite{Djellouli2024ShellMetafluids}.
Even a Newtonian fluid, when perturbed with chaotic surface waves, can be tuned to exhibit arbitrary combinations of drag and diffusion coefficients \cite{Welch2016FluidsTunable}.
Other metafluids can feature energy storage, odd viscosity, multistability, and other emerging properties. 
The applicability of these metafluids has yet to be realized in biomedicine, but likely areas of interest include artificial blood, cerebrospinal fluid, and high-contrast fluids for ultrasound imaging and radiology.
    
\subsection{Cooperative biosensors and diagnostic devices}
\label{subsec:diagnosticDevices}

These materials [\S\ref{subsec:BioinspiredMaterials} - \ref{subsec:Metafluids}] are particularly suitable for high-precision detection and biomedical diagnostics.
For example, active probes can measure viscoelastic properties within single cells using microrheology [Fig.~\ref{fig:Biomedical}B], allowing differentiation between physiological and pathological conditions \cite{Zia2018ActiveSimulation}.
Moreover, self-propelled nanoparticles are now being introduced for biosensing [Fig.~\ref{fig:Biomedical}A], allowing to probe multiple microsamples as the sensor moves \cite{Li2017Micro/nanorobotsDetoxification}. 
This detection of specific biochemical compounds on the fly could make adaptive drug delivery and microsurgery [\S\ref{subsec:MicroroboticSwarms}] another step closer to reality \cite{Pacheco2019Self-propelledReview}.

More traditional healthcare diagnostic devices could also be equipped with active matter components to enhance their functionality and replace external laboratory tests.
Artificial cilia [\S\ref{subsec:BioinspiredMaterials}], for instance, could transport blood into a microfluidic point-of-care (PoC) device and mix this biological sample across an array of biosensors for specific analyte detection \cite{Boukherroub2024TheTests}.
The transduction and amplification of these signals to measurable electical currents can be enhanced using a self-assembled [\S\ref{subsec:SelfAssembly}] film of gold nanoparticles, which aggregate or disaggregate depending on the presence of the target analyte in the analyzed fluid. 

A key shortcoming in current diagnostic devices is sensor fouling \cite{Boukherroub2024TheTests}, often due to bacterial biofilms [\S\ref{subsec:Biofilms}], which drastically reduces the sensing capabilities.
A potential solution is to use rectification techniques to direct their motility or to mislead these bacteria by manipulating their navigation mechanisms [\S\ref{subsec:Navigation}].
Another option is to use actively self-cleaning materials or self-propelled particles that generate flows to clean the detector sites.

\subsection{Organoids and organs on a chip}
\label{subsec:Organoids}

Organoids and organ-on-a-chip (OoC) technologies represent a transformative leap in biomedical research \cite{Dahl-Jensen2017TheOrganogenesis}, enabling the study of emergent behaviors that arise from interactions among multiple cellular agents, rather than isolated individual cells.
Traditional 2D cell cultures fail to capture the dynamics and system-level properties of human tissues, such as synchronized neural activity, metabolic coupling between organs, and coordinated immune responses. 
This gap can be bridged by integrating organoids with OoC platforms, known as organoids-on-chips, which create controlled microenvironments that mimic physiological conditions \cite{Zhao2024IntegratingDevices}. These platforms allow the study of how tissue-level functions emerge, how they break down due to disease, and these functions can be restored through therapeutic interventions.

Recent advancements have led to several key developments in organoids-on-chips: Vascularized [\S\ref{subsec:cardiovascular}] organoids-on-chips now incorporate perfusable microvasculature, improving oxygen and nitrogen delivery while preventing core necrosis. Multi-organ chips can replicate inter-organ communication [\S\ref{subsec:transduction}], enabling the study of systemic interactions, such as metabolic coupling between the liver and pancreas [Fig.~\ref{fig:Biomedical}E]. Tumor microenvironment models integrate stromal, immune, and endothelial cells, providing a dynamic platform to investigate cancer progression [\S\ref{subsec:CellMigration}] and immune response dynamics \cite{Wang2023AdvancesResearch}. These innovations enhance the physiological relevance of in vitro models, bringing them closer to mimicking real human tissues.

Despite these advances, challenges remain, including the need for standardization, enhanced vascularization, and better integration with numerical simulations to predict emergent behaviors and system failures. Looking ahead, multi-scale modeling, synthetic biology approaches, and AI-driven simulations \cite{Cichos2020MachineMatter} will play critical roles in optimizing these platforms \cite{Wang2023AdvancesResearch}. 
The future vision for organoids-on-chip technologies includes fully integrated patient-specific disease models, automated drug screening platforms, and next-generation regenerative therapies, paving the way for precision medicine. An exciting recent example is the use of brain organoids [Fig.~\ref{fig:Biomedical}F] to understand the physics of brain folding \cite{Karzbrun2018HumanFolding}.
With such interdisciplinary collaborations, these systems have the potential to redefine experimental biology, accelerate drug discovery, and revolutionize personalized medicine, making feasible the transition from research tools to clinical applications.

    
\section{DISCUSSION}
\label{sec:Discussion}

In this review, we presented an overview of biomedical active matter, a young discipline with plenty of room for growth: 
Because living systems span over ten orders of magnitude [Fig.~\ref{fig:SelfOrganization}], some say ``this field has more research questions than scientists to answer them.'' 

However, it is essential for us to stay connected and tackle hard problems together, much in the spirit of collective functionalities.
A key challenge for the coming years will be to unify our subfields: There are countless opportunities to join our knowledge, such as combining metamaterials with diagnostics, biosensors with flow networks, and protein engineering with drug delivery. 
Indeed, linking any two subsections in this review could form an interesting new project. 
But again, it is important to focus on global challenges.

Moreover, despite the potential for breakthroughs, collaborations between fundamental biophysics research and translational medicine remain rare.
To overcome this, people will have to leave their comfort zone, from individual scientists to funding agencies, university administrators, and policy makers.
It will also require more training opportunities \cite{Parthasarathy2015TheCourse} for early-career scientists to develop proficiency in theories and experimental methods across different fields. 
In return, because biomedical active matter welcomes people from effectively all academic backgrounds, including the social sciences and humanities, it has the potential to become one of the most inclusive disciplines.
Together, we can do more than alone.


    \begin{summary}[SUMMARY POINTS]
    \begin{enumerate}
    \item Collective functionalities underpin all living systems, from the subcellular to the multiorganismic scale.
    \item Diseases caused by the breakdown of these collective phenomena can be predicted and detected with models based on fundamendal physics principles.
    \item Novel treatment strategies and biomedical devices are being developed based on bioinspired active materials.
    \item Active nanoparticles are already at the forefront of drug delivery. 
    \end{enumerate}
    \end{summary}

    \begin{issues}[FUTURE ISSUES]
    \begin{enumerate}
    \item The field of active matter is maturing, but most of its connections with biomedine are still in their infancy.
    \item In addition to collective motion, more research should focus on other biological functionalities including collective sensing, signalling, and learning.
    \item While papers often show good agreement between experiments and theory, it is necessary to validate data and models quantitatively across the literature to establish robust therapeutic procedures.
    \item Funding agencies should establish more programs that connect basic research in biophysics with translational medicine.
    \end{enumerate}
    \end{issues}


    \section*{DISCLOSURE STATEMENT}
    The authors are not aware of any affiliations, memberships, funding, or financial holdings that might be perceived as affecting the objectivity of this review. 

    \section*{ACKNOWLEDGMENTS}
    This review was written collectively by all members of the Mathijssen lab over the past two years, during more than 100 group meetings filled with lively discussion.
    We are immensely grateful to our friends, family, and colleagues for their support and feedback on this paper. 
    We thank the NIAID NIH BIOART Source for publicly available biomedical illustrations. 
    A.J.T.M.M. acknowledges funding from the Charles E. Kaufman Foundation (Early Investigator Research Award, KA2022-129523; and New Initiative Research Award KA2024-144001), the National Science Foundation (UPenn MRSEC, DMR-2309043), the University of Pennsylvania (CURF, VIPER, Vagelos MLS, and FERBS programs), and the Research Corporation for Science Advancement (Cottrell Scholar Award CS-CSA-2026-125).
    A.T. acknowledges support from the Simons Foundation (Math+X Grant). M.L. is grateful to the Fulbright Commission for the  Fulbright Senior Award to spend a sabbatical in the Mathijssen lab. This work was supported by the National Science Centre of Poland Sonata Bis grant no. 2023/50/E/ST3/00465 to M.L. 


    \printbibliography

\end{document}